\documentclass[epj]{svjour}
\usepackage{graphicx}

\begin{document}

\title{
Self-organized model for information spread in financial markets}
\author{Zhi-Feng Huang}
\thanks{\emph{E-mail address:} zfh@thp.uni-koeln.de}
\institute{Institute for Theoretical Physics, Cologne University,
D-50923, K\"oln, Germany}
\date{}
\abstract{
A self-organized model with social percolation process is proposed
to describe the propagations of information for different trading
ways across a social system and the automatic formation of various 
groups within market traders. Based on the market structure of this 
model, some stylized observations of real market can be reproduced, 
including the slow decay of volatility correlations, and the fat tail 
distribution of price returns which is found to cross over to an 
exponential-type asymptotic decay in different dimensional systems.
}
\maketitle
Recently, microscopic models for financial markets have attracted 
more and more interests 
\cite{stigler,bp-book,ms-book,farmer,lls,c-b,e-z,lux-m}.
These models are based on the empirical
findings of high-frequency market data, such as the fat tails in the
distribution of price changes, the fast decaying of linear correlation,
as well as the existence of long-term volatility correlations 
\cite{bp-book,ms-book,farmer,m-s},
and the intrinsic structure of financial markets, including the 
mutual interactions among market participants through herding and 
imitation behaviors \cite{c-b,e-z} or switching from one strategy 
group to another \cite{lux-m}.

Although different mechanisms are used in these microscopic models to 
simulate the price formation processes, some of the stylized 
observations of real markets can be reproduced. In one set of these 
models, the Cond-Bouchaud herding model \cite{c-b} and the related 
percolation models \cite{s-s,s-j}, 
the power-law asymptotic behavior in the tail of price
return has been obtained, with an exponent well outside the stable 
L\'evy regime \cite{m-s} as found in real data \cite{lux,gpams}. Recently,
this Cond-Bouchaud percolation model has been modified by introducing
a feedback mechanism between price return $R$ and trader activity 
$a$: $a\rightarrow a+\alpha R$, where $\alpha$ is the factor 
representing the sensitivity to price fluctuations \cite{s-j}.
Then the volatility clustering can be produced, and more interestingly,
the empirically observed asymmetric pattern of sharp peaks and flat
troughs in the market prices \cite{r-s} can be incorporated.

In the percolation-type model \cite{c-b,s-s,s-j}, investors are 
randomly distributed on a lattice with certain concentration and
through the neighboring connections they form the percolation clusters,
corresponding to herding groups or companies with probability $a$ of
trading (activity) which is set to be the same for all groups. However,
the concentration of investors on the lattice is to be pre-setted
for obtaining resonable results. Here we introduce a self-organized
model for information spread based on the social percolation process
\cite{solomon,huang}, where the investor groups with various 
trading activities and sizes are formed automatically as a result of the
propagations of trading information across a social network.

Consider a $d$ dimensional lattice representing a social system with
each site located by a trader, that is, the whole lattice is occupied. 
When participating in the market each trader
obeys a certain trading way, which can be simply represented by the
trading activity measuring the frequency of putting the buy or sell
orders, and she may change her behavior due to the influence of outside
information about trading ways if she believes that the information 
is attractive enough and the new trading way can lead to more profits. 
Suppose that the information of various trading ways 
is not public, and spreads among traders according
to the communication structure of the social system, which for
simplification, is considered as the transfer along nearest-neighbor
links of lattice. Thus, a higher dimensional lattice corresponds to
the society with more connections among agents \cite{huang}.
All the traders accepting the same information form a cluster or group 
and will exhibit the same behavior in the market as they obey the same
trading way. 

First, we study the process of information spread for different trading
ways, characterized simply by activity $a(\tau)$ at each spread step
$\tau$ (and the sensitivity to price change $\alpha (\tau)$ if 
considering the feedback effect as described above \cite{s-j}). In the
way of larger $a(\tau)$ the trading will be made more frequently. Each
information of trading way has an attractiveness $q(\tau)$, related to
trader's personal opinion about its profitability.
The dynamics of the model is:

\begin{itemize}
\item At one step $\tau$, a new information of trading way with 
attractiveness $q(\tau)$ and activity $a(\tau)$ both determined
randomly appears and starts at few informed sites to spread by
nearest-neighbor connections among the 
traders with personal preferences $p_i$ (both the values of $q$ and 
$p_i$ are set between $0$ and $1$ and $\delta p$ shown below is
a small fixed parameter).

\item Each step $\tau$ consists of consecutive substeps of the spread
between traders: After receiving the information from one of her 
neighbors at one substep, trader $i$ will accept it and obey this
new trading way if its attractiveness is larger than
her personal preference, i.e., $q(\tau)>p_i$. Then she will spread the
information further to the other neighbors, and increase her preference 
by $\delta p$, i.e., $p_i \rightarrow p_i +\delta p$, as it
should be more difficult and needs larger attractiveness for someone
to change her mind at the next step $\tau +1$ to another new way. 
Otherwise, the trader $i$ will keep the old state, including the old 
trading way and preference, and will not spread the message to her 
neighbors until the end of this step $\tau$. Then, the next substep
begins with the similar judgement and spread procedure.

\item The spread of one step $\tau$ will continue until there is no 
source of information for propagation and the procedure stops by itself, 
that is, all the neighbors of all the traders accepting the information
have already received it, and a trader group or cluster with members 
accepting the common information is formed. Then the next step $\tau+1$ 
starts with the incoming of another new information of attractiveness 
$q(\tau +1)$ and activity $a(\tau +1)$, also selected randomly
between $0$ and $1$, and has 
the spread procedure similar to above. In particular, for a trader $i$ 
belonging to a cluster of previous information, when she receives the new
information with $q(\tau +1)>p_i$ she will accept it and change her
trading way to the new one. Consequently, the size of that old cluster
will decrease by $1$ with the expansion of a new group. Otherwise, the
trader will still remain in the old cluster. Thus,
if the new information is attractive enough, the spread of it will form
a new group which may invade the old ones or even make some of them 
disappearing. Up to now, all traders within the same cluster share the
same activity $a(\tau)$ which was valid when that cluster was formed.
\end{itemize}

Compared with the social percolation model \cite{solomon,huang}, here 
there is no feedback on the attractiveness 
$q$ of information, instead, new
information with randomly selected attractiveness appears at each new step. 
Moreover, the preferences of investors not accepting the new information
or uninformed are not changed, instead of decreasing, and then the trader
preferences increase with the acceptance of more and more information. 
After a large number of steps for the spread of different trading 
information, the trader preferences of the system are close to the upper 
limit $1$ and the newly appearing information has little influence on the 
trader structure. Thus, through a self-organized process the market 
traders in this social system automatically form groups or clusters of 
various sizes and activities due to the acceptance of different trading
information, and these clusters may correspond to the herding groups of 
investors who imitate each other or individual groups with members using 
the same trading activity $a$ and acting in the same manner. 
Here the groups or clusters with larger size correspond to the agents 
in the real market who have bigger influence, and different activities
of groups represent the phenomenon of the real market that the agents
have different trading timescale and some of them trade more frequently
than the other.

\begin{figure*}
\centerline{
\resizebox{0.65\textwidth}{!}{%
  \includegraphics[angle=-90]{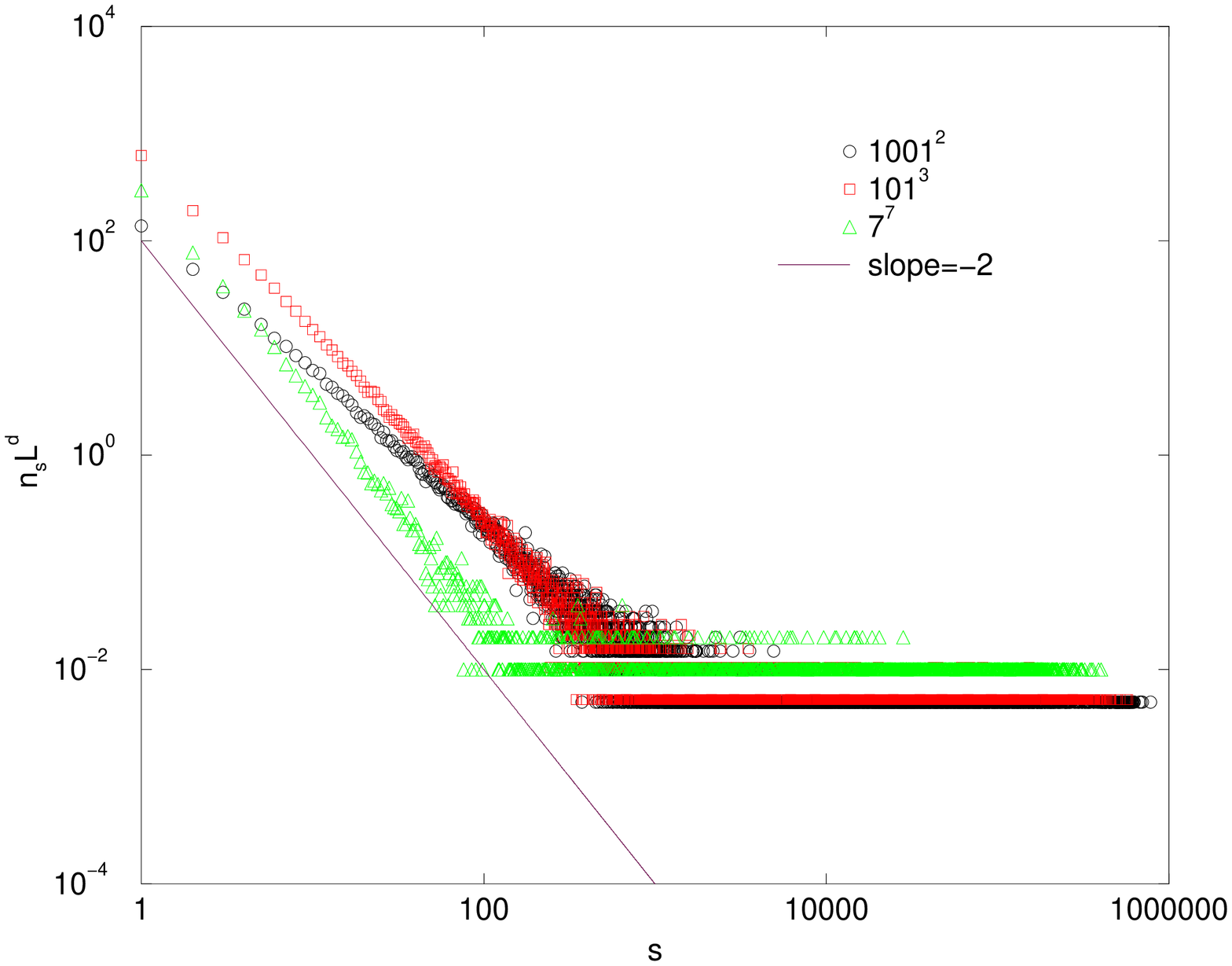}}}
\caption{Log-log plot of the cluster numbers containing $s$ sites after
$\tau=10^5$ steps of the spread model with $\delta p=0.001$, for
$1001^2$ square (averaged over 230 lattices), $101^3$ simple cubic (190 
lattices), and $7^7$ hypercubic (100 lattices) systems. The slope of the 
straight line is $-2$.}
\label{fig-ns}
\end{figure*}

Fig. \ref{fig-ns} shows the results of our simulations for different
hypercubic lattices with length $L$ and dimension $d$, where $n_s$ denotes
the average number of clusters containing $s$ sites, and only two 
parameters are to be chosen: $\delta p=0.001$ and $\tau=10^5$ steps. 
Initially the trader
preferences $p_i$ are distributed randomly between $0$ and $1$, and each
step starts with one randomly selected site on one boundary. Thus, the
sites close to this boundary represent the more informed traders in the
market. A Leath-type algorithm \cite{evertz} is used for the spread
procedures across the lattices with helical boundary conditions.
From Fig. \ref{fig-ns} a power law behavior similar to the percolation 
structure is obtained, but only for small and intermediate sizes of 
clusters. Compared with percolation theory \cite{s-book}, the 
effective power law exponent in the present model is smaller
and can be lower than $2$ in low dimensions with a finite value of 
$\sum sn_s$ which is very close to $1$.
This phenomenon for the slower decaying of $n_s$ may be attributed to 
the much heavier tail in the distribution of cluster sizes in present model. 
In the percolation lattice a cluster with
size roughly comparable to the total system size (largest or "infinite" 
percolating cluster) appears at or above the percolation threshold, while 
in the model here an earlier percolating cluster may be invaded by 
new-coming information that is attractive enough compared with traders' 
preferences, and then its size is diminished with the formation of more
and smaller clusters. Thus, in the present model the sites belonging to
the same cluster need not be nearest-neighbors to each other as in
percolation models \cite{s-s}. For large $s$, it is not clear whether an
asymptotic stretched exponential behavior similar to percolation
structure above the critical threshold exists in present model, and
more detailed work is needed.

Next, we study the trading behaviors of these market groups or clusters
that are determined after enough steps of the spreading process as 
described above. For simplicity, during the trading and price formation 
processes simulated below, the influence of new-coming information, 
i.e., the change of market structure for trader groups, is not considered.
The trading of each group $j$ is represented by an activity $a_j$ obtained 
from different information in the spread process and randomly distributed 
between two limits $a_{\rm max}$ and $a_{\rm min}$, and then at each time 
step, group $j$ decides to make orders with probability $a_j$ (with
equal probability $a_j/2$ for buying or selling) or be inactive with
probability $1-a_j$. This mechanism does not require that each buy order 
should match a sell order, and the balance can be made by assuming the 
involvement of market makers outside the model. Assume that all investors 
in one group contribute the same amount of trading orders, and then the 
trading amount of that group is proportional to the group size.
Moreover, the price return $R$ at each time step $t$
(used for trading process and different from the spread step $\tau$), 
which is defined as the change over time interval $\Delta t$ of 
logarithm of price, is supposed to be proportional to the excess
demand \cite{c-b,farmer2}, that is, the difference between demand and 
supply obtained from the sum over all active clusters,
and normalized by the system size $L^d$, that is, $R\rightarrow R/L^d$,
to avoid the influence of lattice size.

\begin{figure*}
\centerline{
\resizebox{0.65\textwidth}{!}{%
  \includegraphics[angle=-90]{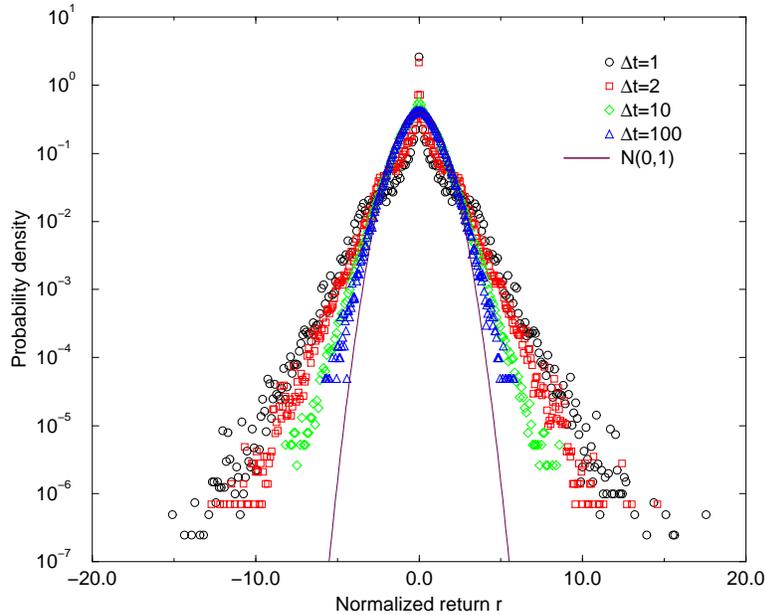}}}
\caption{Semi-log plot of the probability distribution of the normalized
returns for different time intervals $\Delta t=1$, $2$, $10$, and $100$
on $101^3$ cubic lattice with $a_{\rm max}=0.02$ and $a_{\rm min}=0.0001$
(averaged over 190 lattices). A crossover toward the Gaussian distribution
(solid line) is shown with the increasing of time interval.}
\label{fig-3dt}
\end{figure*}

\begin{figure*}
\centerline{
\resizebox{0.65\textwidth}{!}{%
  \includegraphics[angle=-90]{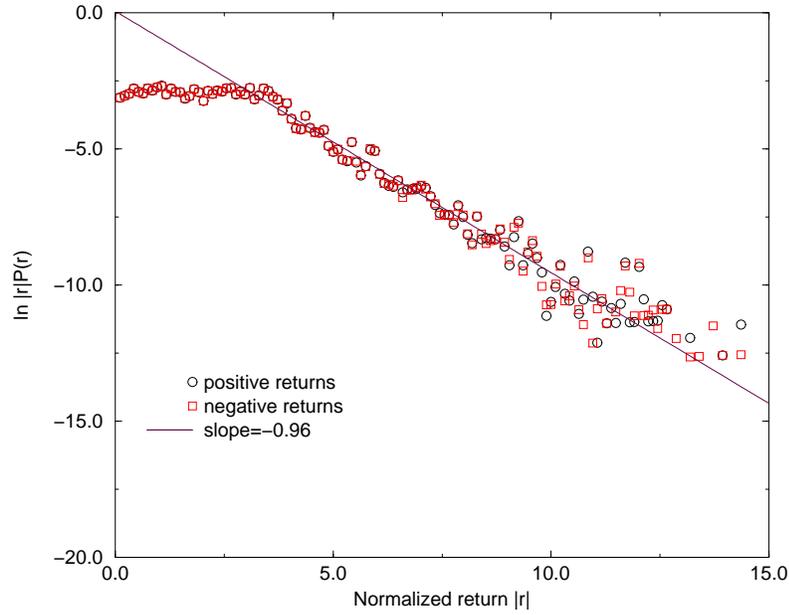}}}
\caption{Replot of Fig. \ref{fig-3dt} for time interval $\Delta t=1$
by showing $\ln |r|P(r)$ as a function of the absolute normalized 
return $|r|$, with a straight line of slope $-0.96$.}
\label{fig-3aexp}
\end{figure*}

Figs. \ref{fig-3dt} and \ref{fig-3aexp} give the results for the
probability distribution of price returns on the simple cubic lattice,
and similar behaviors are found for other dimensions
from two to seven. In the simulations, the unit 
of time increment $\Delta t$ is to be chosen appropriately in order to 
avoid the fluctuations due to random decisions, and here we select $5$ 
time steps as the unit interval $\Delta t=1$. Fig. \ref{fig-3dt} shows
a semi-log plot of the probability distribution of normalized returns, 
defined as $r=(R-\langle R \rangle)/\sigma$ with the average of return 
$\langle R \rangle$ (about $0$) over the time series and the volatility 
$\sigma=(\langle R^2 \rangle -\langle R \rangle ^2)^{1/2}$,
for $a_{\rm max}=0.02$, $a_{\rm min}=0.0001$, and different 
time scales $\Delta t=1$, $2$, $10$, and $100$. With the increasing 
of time interval $\Delta t$, a crossover toward the Gaussian
distribution is observed from the figure, in agreement with the finding
of empirical financial data \cite{gpams}.

In our simulations the asymptotic form of the fat tail distribution
for small time scales is exponential-type, faster
than the power law behavior with exponent about 4 found in recent
empirical studies \cite{lux,gpams} and theoretical simulations 
\cite{e-z,lux-m,s-s}. Similar phenomena can be found in many real 
systems where the probability distributions often cross over to an
exponential-type decay and exhibit curvature in the log-log plots
after a limited range of scales for power law behavior, and a 
stretched exponential description has been shown to account well for 
many natural and economic distributions \cite{l-s}. Very recently, 
an analytic form for the whole range of probability distribution and 
the corresponding Langevin equation have been derived \cite{t-h} based 
on a theorem of general stationary random process and the Hong Kong
Hang Seng Index data, and the asymptotic behavior for large price 
changes is an exponential-type decay: $P(r)\sim\exp (-\beta |r|)/|r|$.
Fig. \ref{fig-3aexp} is transformed from Fig. \ref{fig-3dt} by plotting 
$\ln |r|P(r)$ as a function of the absolute normalized return $|r|$ 
for the data of time interval $\Delta t=1$. Good fits to straight line 
in the tail region are found with $\beta =0.96 \pm 0.02$ (positive 
tail) and $0.96 \pm 0.03$ (negative tail) (fitted over well 
averaged region).

\begin{figure*}
\centerline{
\resizebox{0.65\textwidth}{!}{%
  \includegraphics[angle=-90]{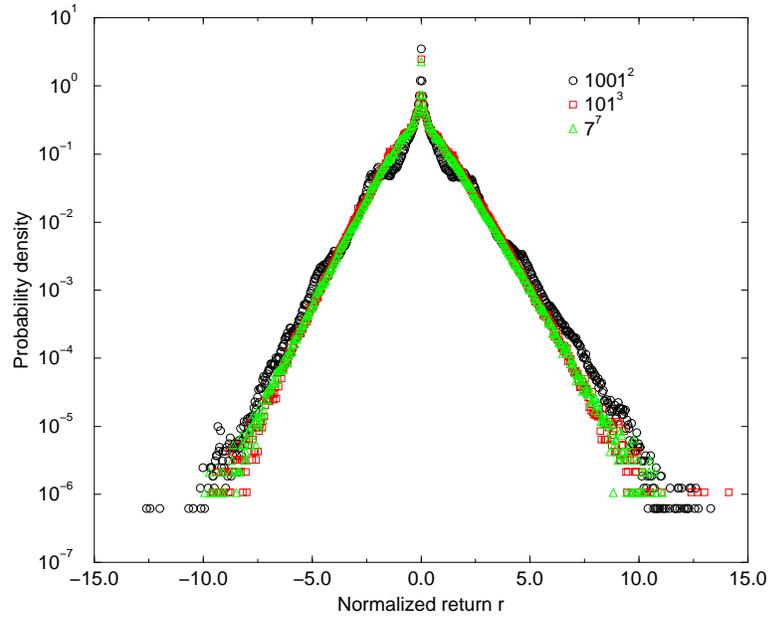}}}
\caption{Semi-log plot of the probability distribution of the normalized
$\Delta t=1$ returns $r$ with $\alpha_{\rm max}=0.1$, 
$\alpha_{\rm min}=0$, $a_{\rm max}=0.2$, and $a_{\rm min}=0.003$, for
$1001^2$ square (200 lattices), $101^3$ cubic (100 lattices), and
$7^7$ hypercubic (100 lattices) systems. The feedback mechanism between
price fluctuation and group activities is considered.}
\label{fig-d237a}
\end{figure*}

\begin{figure*}
\centerline{
\resizebox{0.65\textwidth}{!}{%
  \includegraphics[angle=-90]{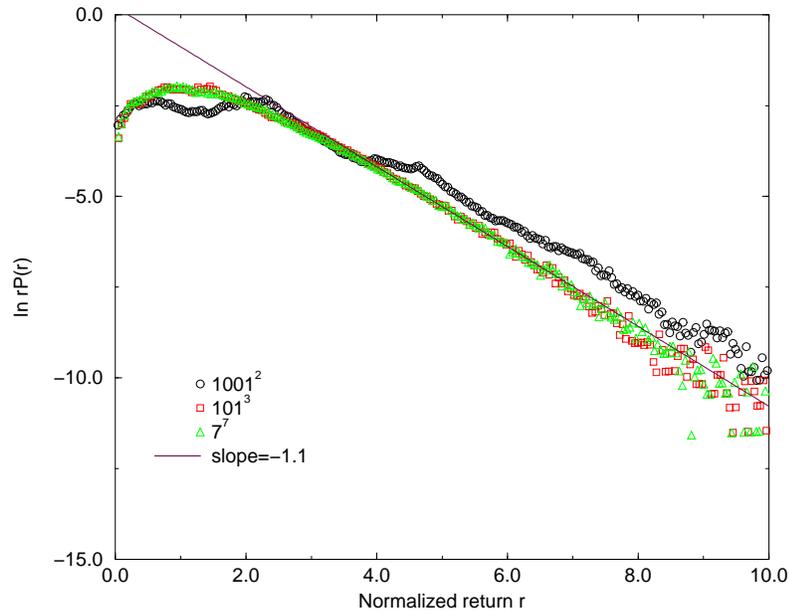}}}
\caption{Replot of Fig. \ref{fig-d237a} for $\ln rP(r)$ as a function of 
the normalized positive return $r$, with a straight line of slope $-1.1$.}
\label{fig-exp}
\end{figure*}

\begin{figure*}
\centerline{
\resizebox{0.65\textwidth}{!}{%
  \includegraphics[angle=-90]{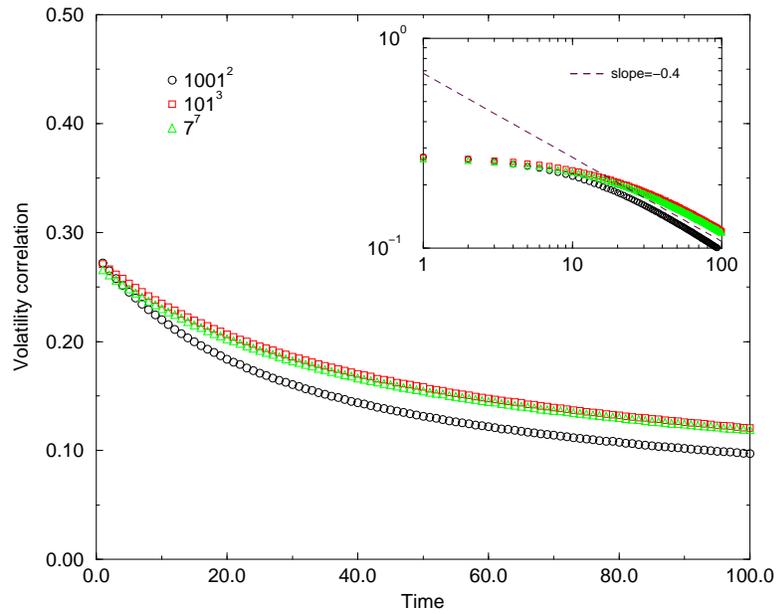}}}
\caption{Autocorrelations for the volatility in different simulations
with the same parameters as Fig. \ref{fig-d237a}. The slow decay is
displayed, and the inset gives the corresponding log-log plots which
exhibit an asymptotic power-law behavior with an exponent about $0.4$
for different systems.}
\label{fig-corr}
\end{figure*}

However, in above simulations the persistence of volatility correlation
is absent, similar to the standard Cond-Bouchaud model \cite{c-b}.
To reproduce the empirical facts of volatility clustering and asymmetry
of bubbles and crashes in stock markets, we follow the method in
\cite{s-j} to introduce a feedback between price fluctuations and 
group behaviors at each time step: $a_j(t+1)=a_j(t)+\alpha_j R$, that is,
the investors are encouraged by the price increasing with more tradings,
and are more prudent when price decreases. We keep $a_j$ in the interval
from $a_{\rm min}$ to $a_{\rm max}$. Furthermore, as the trading
ways of the groups are different with different 
viewpoints on the price changes, in our simulations the parameters
$\alpha_j$ are set randomly between two limits $\alpha_{\rm max}$ and
$\alpha_{\rm min}$. Figs. \ref{fig-d237a} to \ref{fig-corr} show the 
results of simulations in two, three, and seven dimensions, where 
$\alpha_{\rm max}=0.1$, $\alpha_{\rm min}=0$, and the cutoffs of activity 
$a_{\rm max}=0.2$ and $a_{\rm min}=0.003$. To make the 
initial values of group activities irrelevant, the first $10^6$ time 
steps of the simulations are skipped.

The properties of fat tails in the probability distribution of 
normalized price returns $r$ are shown in Fig. \ref{fig-d237a},
for different simulations of time interval $\Delta t=1$ 
in square, cubic, and $d=7$ hypercubic systems, respectively. 
The exponential-type tail behaviors can also be observed in the figure,
and similarly we transformed Fig. \ref{fig-d237a} into
Fig. \ref{fig-exp} by plotting $\ln rP(r)$ vs normalized positive returns 
$r$, with the similar behaviors for negative ones. In the tail 
region good fits to straight line similar to that of the 
simulations without feedback effect are obtained, with parameter value 
$\beta\approx 1.1$ for different dimensional
systems ($\beta =1.05 \pm 0.01$ for size $1001^2$, $\beta=1.108 \pm 0.007$ 
for size $101^3$, and $\beta=1.109 \pm 0.007$ for size $7^7$, 
fitted over well averaged data). Simulations for other dimensions
give the similar results.

\begin{figure*}
\centerline{
\resizebox{0.65\textwidth}{!}{%
  \includegraphics[angle=-90]{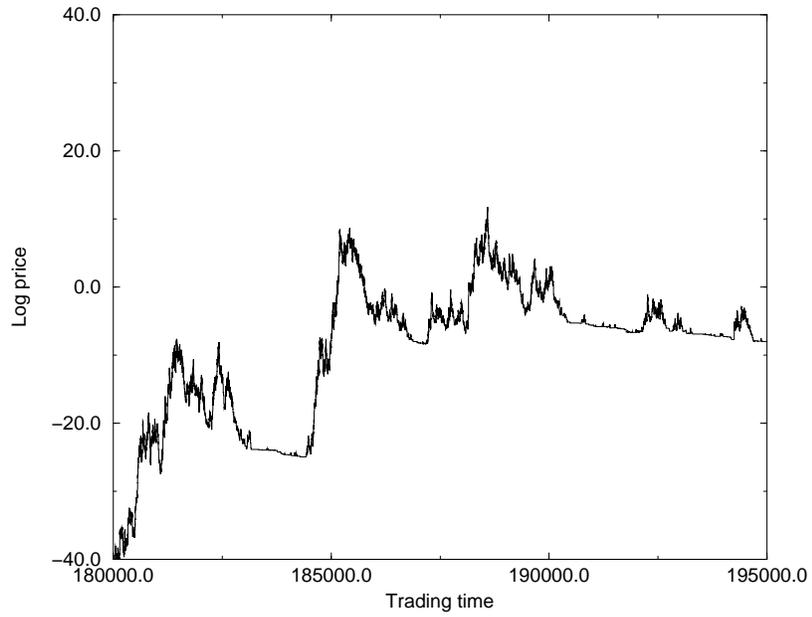}}}
\caption{Time series of the logarithm of price in one simulation with 
the same parameters as Fig. \ref{fig-d237a}, except for 
$a_{\rm min}=0.0003$, on a $31^3$ cubic lattice.}
\label{fig-price}
\end{figure*}

The persistence of long-range volatility correlation, which is defined 
as the autocorrelation between the absolute value of price return, i.e.,
$[\langle |R(t)||R(t+T)| \rangle-\langle |R(t)|\rangle
\langle |R(t+T)|\rangle]/[\langle |R(t)|^2\rangle-
\langle |R(t)|\rangle^2]$, is shown in Fig. \ref{fig-corr} for
$2$, $3$, and $7$ dimensional lattices and time scale $\Delta t=1$,
while the linear autocorrelation for the price returns is around zero.
The inset of Fig. \ref{fig-corr} gives the slow decay of volatility
correlations in log-log plots, representing an asymptotic power-law
with an exponent about $0.4$, similar to the empirical observation
\cite{ms-book}. Volatility clustering as well as
the asymmetry between sharp peaks and shallow valleys of the prices 
found empirically \cite{r-s} are shown in Fig. \ref{fig-price}, which
gives the time series of one simulation on a $31^3$ simple cubic 
lattice.

Compared with the Cond-Bouchaud percolation models, where the system is
fixed at (or slightly above) the percolation threshold 
or set at varying concentrations with results to be integrated over 
\cite{s-s,s-j}, the model here can automatically generate the market
structure among traders, i.e., the trading groups or clusters,
based on the social percolation process which is intrinsically
different from the Cond-Bouchaud one,
and the largest cluster need not be omitted for obtaining
reasonable results. Moreover, in the present model the trading activity
for different groups are different, reflecting the fact that in real
market the agents have various timescales or frequencies of trading.

The qualitative results of the model here are robust with respect to 
parameters. The simulations with different parameters all lead to the 
same stylized features, including the properties of cluster size 
distribution after large enough steps of information spread, the fat 
tail of price return distribution described by exponential-type 
asymptotic decay, the long-range persistence of volatility 
autocorrelations, and the asymmetry between peaks and troughs of the 
prices. Different dimensional lattices simulate the different social 
societies with various degree of personal connections, and the 
dimensionality of the system does not influence the major results 
of the model.

\begin{acknowledgement}
The author would like to thank Dietrich Stauffer and Sorin Solomon 
for very helpful discussions and comments. This work was supported 
by SFB 341.
\end{acknowledgement}

\end{document}